\documentclass[sigconf,anonymous=false]{acmart}

\AtBeginDocument{%
  \providecommand\BibTeX{{%
    \normalfont B\kern-0.5em{\scshape i\kern-0.25em b}\kern-0.8em\TeX}}}

\setcopyright{acmcopyright}
\copyrightyear{2020}
\acmYear{2020}
\acmConference[CIKM'20]{CIKM'20: International Conference on Information and Knowledge Management}{October 19--23, 2020}{Ireland}
\acmBooktitle{CIKM'20: International Conference on Information and Knowledge Management, October 19--23, 2020, Ireland}
\newcommand{\algname}[1] {{\fontfamily{cmtt}\selectfont {#1}}}
\usepackage{todonotes}
\usepackage{bbm}
\usepackage{amssymb}
\usepackage{dsfont}

\begin{document}

%\title{Feedback Loop and its Impact on Bias Amplification in Recommender Systems}
\title{Feedback Loop and Bias Amplification in Recommender Systems}

\author{Masoud Mansoury}
\authornote{This author also has affiliation in School of Computing, DePaul University, Chicago,
USA, mmansou4@depaul.edu.}
\affiliation{
  \institution{Eindhoven University of Technology}
  \city{Eindhoven}
  \state{the Netherlands}
}
\email{m.mansoury@tue.nl}

\author{Himan Abdollahpouri}
\affiliation{
    \institution{University of Colorado Boulder}
    \city{Boulder}
    \state{USA}
}
\email{himan.abdollahpouri@colorado.edu}

\author{Mykola Pechenizkiy}
\affiliation{%
 \institution{Eindhoven University of Technology}
 \city{Eindhoven}
 \country{the Netherlands}}
\email{m.pechenizkiy@tue.nl}

\author{Bamshad Mobasher}
\affiliation{%
  \institution{DePaul University}
  \city{Chicago}
  \country{USA}}
\email{mobasher@cs.depaul.edu}

\author{Robin Burke}
\affiliation{
  \institution{University of Colorado Boulder}
  \city{Boulder}
  \country{USA}}
\email{robin.burke@colorado.edu}

\renewcommand{\shortauthors}{Masoud Mansoury, et al.}

\begin{abstract}
  Recommendation algorithms are known to suffer from popularity bias; a few popular items are recommended frequently while the majority of other items are ignored. These recommendations are then consumed by the users, their reaction will be logged and added to the system: what is generally known as a \textit{feedback loop}. In this paper, we propose a method for simulating the users interaction with the recommenders in an offline setting and study the impact of feedback loop on the popularity bias amplification of several recommendation algorithms. We then show how this bias amplification leads to several other problems such as declining the aggregate diversity, shifting the representation of users' taste over time and also homogenization of the users experience. In particular, we show that the impact of feedback loop is generally stronger for the users who belong to the minority group.
\end{abstract}




\maketitle

\section{Introduction}\label{intro}
%Recommender systems are powerful tools in generating personalized recommendations for users. There are various recommendation approaches in the literature. Collaborative Filtering (CF) is one of the well-known recommendation approaches that works based on the similarity between users and items. These systems, while effective, are biased against certain groups of users and items. Bias often originates from input data or the recommendation algorithm.   

\begin{figure}
    \centering
    \includegraphics[width=3in]{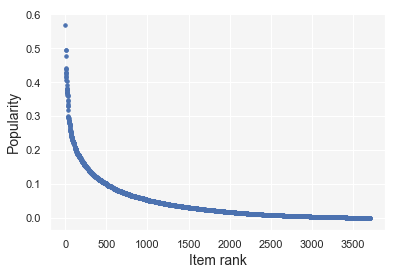}
    \caption{Item popularity in the MovieLens dataset. Items are sorted based on popularity from the most (left) to the least (right): lower \textit{item rank} has higher popularity.}
    \label{fig:longtail}
\end{figure}

Collaborative Filtering (CF) is a well-known recommendation technique that uses historical data on interactions between users and items (i.e., ratings provided by the users on different items), and generates personalized recommendations for the users. Recommendations generated by CF generally suffer from bias against certain groups of users or items \cite{yin2012challenging,yao2017}. Bias in recommendation output can originate from different sources: 1) it may stem from the underlying biases in the input data: Figure ~\ref{fig:longtail} shows the distribution of the rating data in the MovieLens dataset (see section ~\ref{expo} for more details on this dataset) where a few popular items receive large proportion of ratings while the majority of other items do not receive much attention from the users, or 2) it may be due to the algorithmic bias where recommendation algorithms propagate the existing bias in data \cite{dietmar2013} and, in some cases, intensify it by recommending these popular items even to the users who are not interested in popular items \cite{abdollahpouri2020multi}.    

% In the case of users, some user groups may have better representation in the input data (e.g., number of ratings) and this well-representation compared to other users' groups causes recommendation algorithms deliver better recommendations (i.e., more accurate, diverse, calibrated, and so on) to these groups of users \cite{mansoury2020a}. 

The algorithmic bias could be intensified over time when users interact with the given recommendations, that are biased towards popular items, and this interaction is added to the data. Users receiving recommendation lists may select (e.g., by rating or clicking) some of the recommended items and the system will add those items to their profiles as part of their interaction history. In this way, recommendations and user profiles form a feedback loop \cite{damour2020,chaney2018}; the users and the system are in a process of mutual dynamic evolution where users profile get updated over time via recommendations generated by the recommender system and the effectiveness of the recommender system is also affected by the profile of users.        

The study on feedback loop in machine learning and particularly recommender systems has recently received more attention from researchers \cite{damour2020,jiang2019,chaney2018,schmit2017,sinha2016,sun2019debiasing}. D'Amour et al. \cite{damour2020} analyzed the long-term fairness of machine learning based decision-making systems in three different domains through simulation studies: bank loans, allocation of attention, and college admission in an agent-based environment. Their analysis showed that common single-step analysis does not show the dynamic behavior of the system and the need for exploring the long-term effect of the decision-making systems. In another work which is also based on a simulation using synthetic data, Chaney et al. \cite{chaney2018} showed that feedback loop causes homogenization of the user experience and shift in item consumption. Homogenization in their study was measured as the ratio of commonly rated items in a target user's profile and her nearest neighbor's profile, and showed that homogenization leads to lower utility for the users. 

In this paper, we investigate the effect of feedback loop on amplifying bias in recommender systems. We study popularity bias amplification and the impact of this effect on other aspects of a recommender system including declining aggregate diversity, shifting the representation of the users' taste, and also homogenization of the users. In particular, we show that the impact of feedback loop is generally stronger for the users who belong to the minority group. For the experiments, we simulate the users interaction with recommender systems over time in an offline setting. The concept of time here is not chronological but rather consecutive interactions of users with the recommendations in different iterations. That is, in each iteration, users' profile is updated by adding selected items from the recommendation lists generated at previous iteration to their profile. We performed the simulation using three recommendation algorithms on a movie dataset.  

\section{Feedback loop simulation}\label{simulation}

The idealistic scenario for investigating the effect of feedback loop on amplifying bias in recommender systems is to perform online testing on a real-world platform with steady stream of data. 
%In this scenario, it is possible to deliver recommendations to the users and then get their feedback about those recommendations (i.e., whether or not a user clicked on the recommended items) to measure the amount of bias they perceived. 
However, due to the lack of access to the real-world platforms for experimentation, we simulate the recommender system process in an offline setting. To do so, we simulate recommendation process over time by iteratively generating recommendation lists to the users and updating their profile by adding the selected items from those recommendation lists based on an acceptance probability. 
%In our simulation, we translate each iteration as a time point. 
Given the rating data $D$ as an $m \times n$ matrix formed by ratings provided by the users $U=\{u_1,...,u_m\}$ on different items $I=\{i_1,...,i_n\}$, the mechanism for simulating feedback loop is to generate recommendation lists for the users in each iteration $t \in \{1,...,T\}$ and updating their profile based on the delivered recommendations in each iteration. The following steps show this mechanism: 

\begin{itemize}
    \item[\textbf{1)}] Given $D^t$ as the rating data in iteration $t$, we split $D^t$ into training and test sets as 80\% for $train^t$ and 20\% for $test^t$. %Note $S^0$ is the existing data collected from a platform.
    \item[\textbf{2)}] We build the recommendation model on $train^t$ to generate the recommendation lists $R^t$ to all users. %Then, we evaluate the performance of recommendation model and the amount of bias in $R^t$ using $test^t$ at time $t$.
    \item[\textbf{3)}]  For each user $u$ and recommendation list $R_{u}^{t}$ generated for $u$, we follow the \textit{acceptance probability} concept proposed in \cite{abdollahpouri2019beyond} to decide which item from the recommendation list the user might select. The acceptance probability assigns a probability value to each item in $R_{u}^{t}$ where more relevant items (higher ranked) are assigned higher probability to be selected. Formally, for each item $i$ in $R_{u}^{t}$, the acceptance probability can be calculated as follows:
    
    \begin{equation}\label{acceptanceprob}
        prob(i|R_{u}^{t})=e^{\alpha \times rank_i}
    \end{equation}
    
    \noindent where $\alpha$ is a negative value ($\alpha < 0$) for controlling the probability assigned to each recommended item and $rank_i$ is the rank of the item $i$ in $R_{u}^{t}$. 
    Equation \ref{acceptanceprob} is only a selection probability and does not assign a potential rating a user might give to the selected item. This is particularly important if we want to also include rating-based algorithms such as \algname{UserKNN} in our simulation as we have done it in this paper. To estimate the rating a user might give to the selected item, we follow the \textit{Item Response Theory} used in \cite{sinha2016,ho2008}. More formally,
    
     \begin{equation}\label{irt1}
        \omega=\overline{s}_u + (sd(s_u) \times \overline{s}_i) + \eta_{u,i}
    \end{equation}

    \noindent where $\overline{s}_u$ is the average of the ratings in $u$'s profile, $sd(s_u)$ is the standard deviation of the ratings in $u$'s profile, $\overline{s}_i$ is the average of ratings assigned to $i$, and $\eta_{u,i}$ is a noise term derived from a Gaussian distribution (i.e., $\eta_{u,i} \sim N(0,1)$). In order to estimate an integer rating value in the range of $[a,b]$ where $a$ and $b$ are the minimum and maximum rating values, respectively, we use the following equation \cite{sinha2016}:
    
    \begin{equation}\label{irt2}
        \hat{s}_{u,i}=max(min(round(\omega),b),a)
    \end{equation}
    
    After estimating $\hat{s}_{u,i}$, we add $(i,\hat{s}_{u,i})$ to $u$'s profile if $i$ is not already in $u$'s profile and we repeat this process for all users to form $D^{t+1}$. %(updated train set at time $t+1$).
    
\end{itemize}

The steps 1 through 3 are repeated in each iteration.

% With above process, there will be a training set, a test set, and a recommendation set at each iteration $t$ that can be used to calculate various performance metrics and to measure the amount of bias in recommendation results. 

\section{Modeling Feedback}\label{biasmodel}

As we mentioned in section ~\ref{intro}, recommendation algorithms suffer from  popularity bias. In this section, we formally model the propagation of this bias due to the feedback loop phenomenon. Let $\overline{P}_{D^t}$ and $\overline{P}_{R^t}$ be the average popularity (i.e. the expected values) of the items in the rating data and the recommended items in iteration $t$, respectively. 

\begin{equation}\label{modeling}
    \overline{P}_{R^t} \propto \overline{P}_{D^t}+\theta^t 
\end{equation}

\noindent where $\theta^t$ is the percent increase of the popularity of the recommendations compared to that of rating data in iteration $t$. Now, assuming, out of all the recommendations given to the users, we add $K$ interactions ($K>=0$) to the profiles of the users, the size of the rating data in the next iteration would be $|D^t|+K$ and its average popularity will be $\overline{P}_{D^{t+1}} \approx \frac{|D^t|\times \overline{P}_{D^t} + K \times (\overline{P}_{D^t}+\theta^t) }{|D^t|+K}$ which can be simplified as $\frac{(|D^1|+K) \times \overline{P}_{D^t} +K \times \theta^t}{|D^t|+K}=\overline{P}_{D^t}+\frac{K \times \theta^t}{|D^t|+K}$ 
which means the average popularity of the items in the rating data is now increased by $\frac{K \times \theta^t}{|D^t|+K}$. Based on Equation ~\ref{modeling}, by definition, the average popularity of the recommended items in each iteration is proportional to the average popularity of the rating data in the same iteration plus a positive value and since $\overline{P}_{D^{t+1}}$ has increased compared to $\overline{P}_{D^t}$, $\overline{P}_{R^{t+1}}$ will be also higher than $\overline{P}_{R^t}$ due to transitivity. In other words, in each iteration $t$, $\overline{P}_{R^{t+1}}>\overline{P}_{R^{t}}$ indicating the popularity propagation/intensification from one iteration to the next one.  

\section{Experiments}\label{expo}

In this section, we describe the data and the algorithms we used in our experiments along with the empirical results.

\subsection{Data}
We performed our experiments on MovieLens 1M\footnote{We picked this dataset particularly because it has information about both the users
(such as gender) and the items (such as genre).} dataset \cite{harper2015} which is a movie rating data collected by the GroupLens research group. In this dataset, 6,040 users provided 1,000,209 ratings (4,331 males provided 753,769 ratings and 1,709 females provided 246,440 ratings) on 3,706 movies. The ratings are in the range of 1-5 and the density of the dataset is 4.468\%. Also, each movie is assigned either a single genre or a combination of several genres. Overall, there are 18 unique genres in this dataset. 

\subsection{Algorithms}

We performed a comprehensive evaluation of the effect of feedback loop on amplifying bias in recommender systems using three different recommendation algorithms: user-based collaborative filtering (\algname{UserKNN}) \cite{Resnick:1994a}, bayesian personalized ranking (\algname{BPR}) \cite{rendle2009bpr}, and \algname{MostPopular}. \algname{BPR} is a factorization model that works on binary data and \algname{UserKNN} is a neighborhood model that works on explicit rating data. \algname{MostPopular} recommends the most popular items to everyone (the popular items that a user has not seen yet). We set the number of factors in \algname{BPR} and the number of neighbors in \algname{UserKNN} to 50 to achieve the best performance in terms of precision. For our simulation, we performed the steps 1-3 in section \ref{simulation} for 20 iterations ($T=20$).

\begin{figure}[h]
    \centering
    \includegraphics[width=0.47\textwidth]{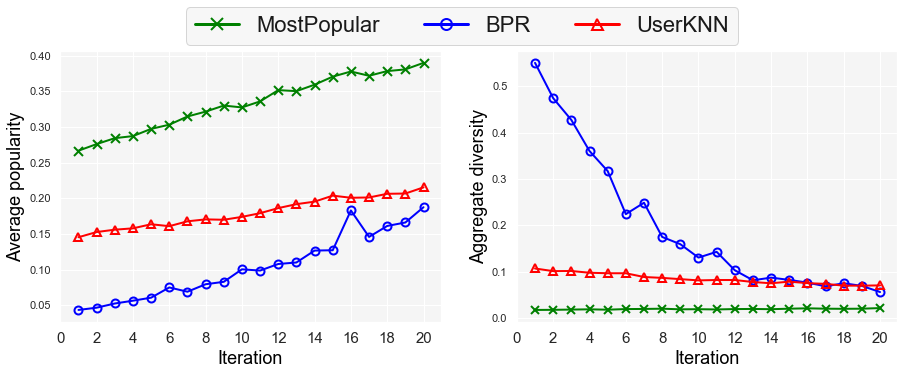}
    \caption{Average popularity (left) and aggregate diversity (right) of the recommendations.}\label{fig:popularity}
\end{figure}

\subsection{Results}

\subsubsection{Popularity bias amplification}

As we formally showed in section \ref{biasmodel}, recommendation models can intensify the popularity bias in input data over time due to the feedback loop. Figure \ref{fig:popularity} (left) shows the effect of such a loop on the average popularity of recommendation lists over time (i.e. in different iterations). As shown in this plot, even though these algorithms start with different average popularity values due to their inherent nature, they all show an ascending pattern in terms of the average popularity over different iterations. The curve for \algname{BPR} seems to have a steeper slope compared to the other algorithms indicating a stronger bias propagation of this algorithm. The exact reason for these performance differences across different algorithms needs further investigation and we leave it for future work. 

% This result is due to the compounding of bias from one iteration to the other iterations: tendency of recommendation models in recommending popular items at iteration $t$ will cause increasing popularity of train data at iteration $t+1$ as more popular items would be selected from the recommendation lists at iteration $t$. In contrast, \algname{Random} is unaffected by popularity bias and even slightly decreases the popularity bias over time, because it does not have any popularity bias and there is no possibility of compounding.

Figure \ref{fig:popularity} (right) shows the aggregate diversity (aka catalog coverage) of recommendation algorithms: the percentage of items which appear at least once in the recommendation lists across all users. As a recommender system concentrates more on popular items, it will necessarily cover fewer items in its recommendations and that effect is clear here, especially for \algname{BPR}, which starts out with a relatively high aggregate diversity.

This bias amplification over different iterations could lead to two other problems: 1) shifting the representation of the user's taste over time, and 2) the domination of one group of users (the majority group) over another (the minority group) which eventually could diminish the differences between the groups and create homogenization.

\subsubsection{Shifting users' taste representation}

One consequence of the feedback loop is shifting the representation of the users' taste revealed in user profiles. We define the users interest toward various movie genres based on the rated items in their profile which creates a genre distribution over rating data. This genre distribution is calculated as the ratio of the movies associated with each genre over different genres in the users' profiles. In the MovieLens dataset, some movies are assigned multiple genres hence, in those case, we assign equal probability to each genre. For example, if an item has genres $a$ and $b$, the probability of either of $a$ and $b$ is 0.5. Given genre distribution in iteration $t=1$ as initial preferences represented in the system, we are interested in investigating how initial users' taste representation changes over time due to the feedback loop. For this purpose, in each iteration $t>1$, we calculate the Kullback-Leibler divergence (KLD) between the initial genre distribution and the genre distribution in iteration $t$ for each user. Higher KLD value indicates higher deviation from the initial preference.  

\begin{figure}[h]
    \centering
    \includegraphics[width=0.48\textwidth]{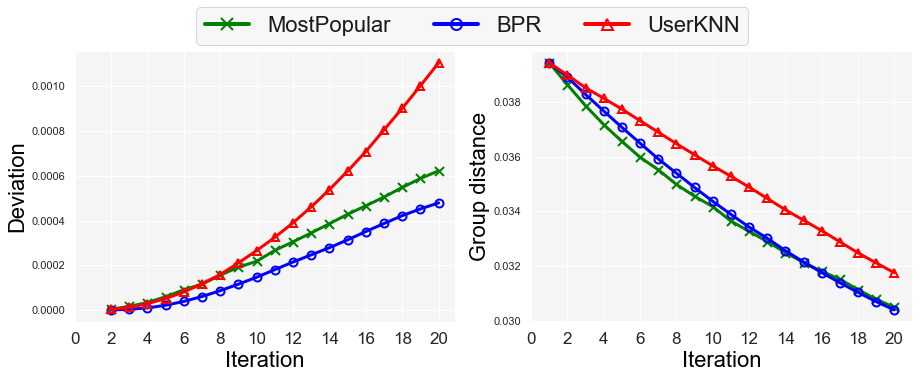}
    \caption{Deviation from the initial preferences (left) and the distance between the representation of genre preferences of males and females in different iterations.}\label{fig:homogenity}
\end{figure}

Figure \ref{fig:homogenity} (left) shows the deviation of users taste from their initial preferences. In all recommendation algorithms, we observe that the deviation of users' profiles from their initial preferences increases over time. It is worth noting that the change in users preferences shown in this Figure is the change in the representation of users' preferences in the system, not the change in users' intrinsic preferences. One consequence of this change in representation of users' preferences in the system is that recommendation models may not be able to capture the users' true preferences when generating recommendations for the users.

\subsubsection{Homogenization}

Shifting the users' taste representation could happen in two situations: when the recommendations given to the users are more diverse from what the users are interested in (i.e. exploration), or when the recommendations are over-concentrated on few items when the users' profiles are more diverse. In the latter, since all users are exposed to a limited number of items over time, their profiles all converge towards a common range of preferences.

Figure \ref{fig:homogenity} (right) shows the distance between the representation of males (majority group) and females (minority group) preferences over time. In each iteration $t$, given the genre distribution separately extracted from males and females ratings as $G_M$ and $G_F$, respectively, we calculate the KLD of $G_M$ and $G_F$, $KLD(G_M||G_F)$, which measures the distance between the preferences of males and females. As shown in the plot, the KLD value dramatically decreases over time in all algorithms showing the strong homogenization of users' preferences.

Now, an interesting question is which user group is dominating the other. To answer this question we separately compare genre preferences of males and females with the preferences of the whole population. Given $G$ as the initial genre preferences of all users (the population), we calculate $KLD(G||G_F)$ and $KLD(G||G_M)$ in each iteration $t$. %we compare the distance between $G$ and $G_M$ with the distance between $G$ and $G_F$.

% Figure \ref{fig:population} (left) shows the preference deviation from $G$ separately for males and females. It can be observed that in all algorithms, the distance between $G$ and $G_F$ decreases over time showing that females preferences are approaching toward the population preferences. Also, the low distance between $G$ and $G_M$ shows that the population is more represented by males as the majority group. Therefore, we can conclude that the females preferences are approaching toward the males preferences over time. Again, it is important to note that this is not the users' real preferences, and is only the representation of the users' preferences in the system.

\begin{figure}[h]
    \centering
    \includegraphics[width=0.49\textwidth]{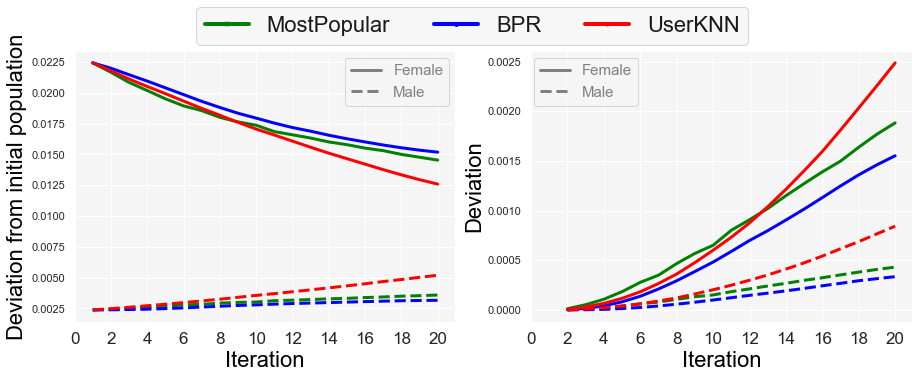}
    \caption{Deviation of the representation of males and females preferences from the representation of the population initial preferences (left) and deviation of the representation of males and females preferences from their initial preferences (right).}\label{fig:population}
\end{figure}

Figure \ref{fig:population} (left) separately shows $KLD(G||G_F)$ and $KLD(G||G_M)$ in different iterations. We can see that, for all algorithms, the representation of females preferences are approaching toward the representation of initial preferences of the population. However, this value is slightly increasing for males showing that they become distant from the preferences of the initial population. We believe the reason is that male users are taking up the majority of the ratings in the data and hence, initially, the population is closer to the male profiles. Over time, since the recommended items are more likely to be those rated by males (as males have rated more items), when added to the users' profiles, causes the female profiles to get closer to the initial population which was dominated by the male users. 

Figure \ref{fig:population} (right) shows the deviation from the representation of initial preferences of each user in the system separately for males and females. In all algorithms, the deviation for females is significantly higher than males, demonstrating the severity of the impact of the feedback loop on the minority group (e.g. females in our experiment). 

% Figure \ref{fig:popularity-deviation} (left) shows the average popularity of recommended items separately for males and females. The plot shows that \algname{MostPopular}, \algname{BPR}, and \algname{UserKNN} amplify the popularity bias for female users more than male users. In particular, \algname{BPR} starts with small difference in average popularity between males and females, but substantially increases this difference over time. One reason for this difference between males and females is that males are majority in this dataset meaning that there are more male users than female users and also there are more ratings provided by males than females. The same patterns can also be observed in terms of aggregate diversity.

\section{Discussion and Future Work}

There are several interesting threads of research that could be built on this work. Firstly, in some recommendation domains such as music, it is very common for a user to listen to the same song repeatedly. Therefore, the restriction we imposed on the selection algorithm in this paper regrading the items that were already in the users' profile (those items were not added to the users' profiles in the next iteration) could be lifted and, instead, the rating for that item would be updated in each iteration.  

Secondly, different strategies for user grouping could be used. Here, we used a pre-defined label for users (i.e. gender) to create user groups. One could group the users based on their average profile size, average popularity of their rated items, or some other statistical characteristics that might be of importance for any particular reason.  

Thirdly, different algorithms that control the popularity bias problem \cite{antikacioglu2017,mansoury2020b,kamishima2014correcting} could be investigated in terms of how they mitigate the bias amplification in feedback loop. Our hypothesis is that since these algorithms reduce the popularity bias in each iteration, according to Equation ~\ref{modeling} their bias amplification over time would be also smaller than the standard algorithms. 
 
Finally, the selection technique in Equation ~\ref{acceptanceprob} we used in this paper leverages the ranking position of the items in the list in order to define whether it would be selected by the user or not. Other selection policies such as top-1 (selecting the first item in the list) or even random selection could be studied. 

\section{Conclusion}

In this paper, we investigated the effect of feedback loop on bias amplification in recommender systems through an offline simulation. We formally and empirically showed that different recommendation algorithms amplify the existing bias through different iterations of users interaction. We then showed that this bias amplification leads to other issues in recommender systems such as declining the aggregate diversity, shifting the representation of the users preferences (i.e. their profiles), and homogenization of the user groups. In particular, for two user groups males and females, we observed that the bias amplification for the females which happen to be in minority group based on their population and their number of ratings was stronger than that of males. These results emphasize the importance of the algorithmic solutions to tackle popularity bias and increasing diversity in the recommendations since even a small bias in the current state of a recommender system could be greatly amplified over time if it is not addressed properly.

%In future work, we plan to study the effectiveness of various bias mitigation techniques in reducing the bias in recommender systems over time.

\bibliographystyle{ACM-Reference-Format}
\bibliography{ref}

\end{document}